\documentclass[preprint,superscriptaddress]{revtex4-1}
\usepackage{graphicx}
\pdfoutput=1
\begin{document}

\title{Bright nanoscale source of deterministic entangled photon pairs violating Bell's inequality } 





\date{\today}

\author{Klaus~D.~J\"ons}
\email[Corresponding author: ]{klausj@kth.se}
\altaffiliation{This author contributed equally to this work.}
\author{Lucas~Schweickert}
\altaffiliation{This author contributed equally to this work.}
\affiliation{Applied Physics Department, Royal Institute of Technology, Albanova University Centre, Roslagstullsbacken 21, 106 91 Stockholm, Sweden}
\affiliation{Kavli Institute of Nanoscience, Delft University of Technology, Lorentzweg 1, 2628CJ Delft, The Netherlands}
\author{Marijn~A.~M.~Versteegh}
\affiliation{Applied Physics Department, Royal Institute of Technology, Albanova University Centre, Roslagstullsbacken 21, 106 91 Stockholm, Sweden}
\affiliation{Kavli Institute of Nanoscience, Delft University of Technology, Lorentzweg 1, 2628CJ Delft, The Netherlands}
\affiliation{Quantum optics, Quantum Nanophysics and Quantum Information, Faculty of Physics, University of Vienna, Boltzmanngasse 5, 1090 Vienna, Austria}
\affiliation{Institute for Quantum Optics and Quantum Information, Austrian Academy of Science, Boltzmanngasse 3, 1090 Vienna, Austria}
\author{Dan~Dalacu}
\author{Philip~J.~Poole}
\affiliation{National Research Council of Canada, Ottawa, Canada, K1A 0R6}
\author{Angelo~Gulinatti}
\affiliation{Politecnico di Milano, Dipartimento di Elettronica Informazione e Bioingegneria, piazza Leonardo da Vinci 32 - 20133 Milano, Italy}
\author{Andrea~Giudice}
\affiliation{Micro Photon Devices, via Stradivari 4 - 39100 Bolzano, Italy}
\author{Val~Zwiller}
\affiliation{Applied Physics Department, Royal Institute of Technology, Albanova University Centre, Roslagstullsbacken 21, 106 91 Stockholm, Sweden}
\affiliation{Kavli Institute of Nanoscience, Delft University of Technology, Lorentzweg 1, 2628CJ Delft, The Netherlands}
\author{Michael~E.~Reimer}
\affiliation{Kavli Institute of Nanoscience, Delft University of Technology, Lorentzweg 1, 2628CJ Delft, The Netherlands}
\affiliation{Institute for Quantum Computing and Department of Electrical \& Computer Engineering, University of Waterloo, Waterloo, N2L 3G1, Canada}



\begin{abstract}
Global, secure quantum channels will require efficient distribution of entangled photons. Long distance, low-loss interconnects can only be realized using photons as quantum information carriers. However, a quantum light source combining both high qubit fidelity and on-demand bright emission has proven elusive. Here, we show a bright photonic nanostructure generating polarization-entangled photon-pairs that strongly violates Bell's inequality. A highly symmetric InAsP quantum dot generating entangled photons is encapsulated in a tapered nanowire waveguide to ensure directional emission and efficient light extraction. We collect $\sim$200 kHz entangled photon-pairs at the first lens under 80\,MHz pulsed excitation, which is a 20 times enhancement as compared to a bare quantum dot without a photonic nanostructure. The performed Bell test using the Clauser-Horne-Shimony-Holt inequality reveals a clear violation ($S_{\text{CHSH}}>2$) by up to 9.3 standard deviations. By using a novel quasi-resonant excitation scheme at the wurtzite InP nanowire resonance to reduce multi-photon emission, the entanglement fidelity ($F=0.817\,\pm\,0.002$) is further enhanced without temporal post-selection, allowing for the violation of Bell's inequality in the rectilinear-circular basis by 25 standard deviations. 
Our results on nanowire-based quantum light sources highlight their potential application in secure data communication utilizing measurement-device-independent quantum key distribution and quantum repeater protocols.

\end{abstract}

\maketitle

\section*{Introduction}
Quantum light sources providing strongly entangled photons are essential components for quantum information processing~\cite{Bouwmeester.Ekert.ea:2000,Gisin.Ribordy.ea:2002}. In particular, secure quantum communication schemes~\cite{Acin.Brunner.ea:2007,Scarani.Bechmann-Pasquinucci.ea:2009} rely on light sources that meet stringent requirements. Quantum cryptography based on entangled photon-pair sources requires a violation of Bell's inequality to guarantee secure data transfer~\cite{Ekert:1991}. Optically active semiconductor quantum dots satisfy several key requirements for these applications, namely: fast repetition rate up to 2\,GHz~\cite{Hargart.Kessler.ea:2013}, on-demand generation~\cite{Muller.Bounouar.ea:2014}, electrical operation~\cite{Yuan.Kardynal.ea:2002,Salter.Stevenson.ea:2010}, position control at the nano-scale~\cite{Baier.Pelucchi.ea:2004}, and entangled photon pair emission~\cite{Akopian.Lindner.ea:2006,Young.Stevenson.ea:2006}.
However, efficient photon extraction from these quantum dots requires additional photonic structures to be engineered around them, such as micro-cavities~\cite{Moreau.Robert.ea:2001} or nanowire waveguides~\cite{Claudon.Bleuse.ea:2010}, to steer the light emission efficiently in the desired direction. Even though the generation of entangled photons was already reported from such structures~\cite{Dousse.Suffczynski.ea:2010,Versteegh.Reimer.ea:2014,Huber.Predojevic.ea:2014}, a violation of Bell's inequality in micro-cavity structures proved to be elusive due to multi-photon emission and cavity induced dephasing that resulted in a significant reduction of the source fidelity. 

\begin{figure*}[t]
\begin{centering}
\includegraphics[width=0.9\linewidth]{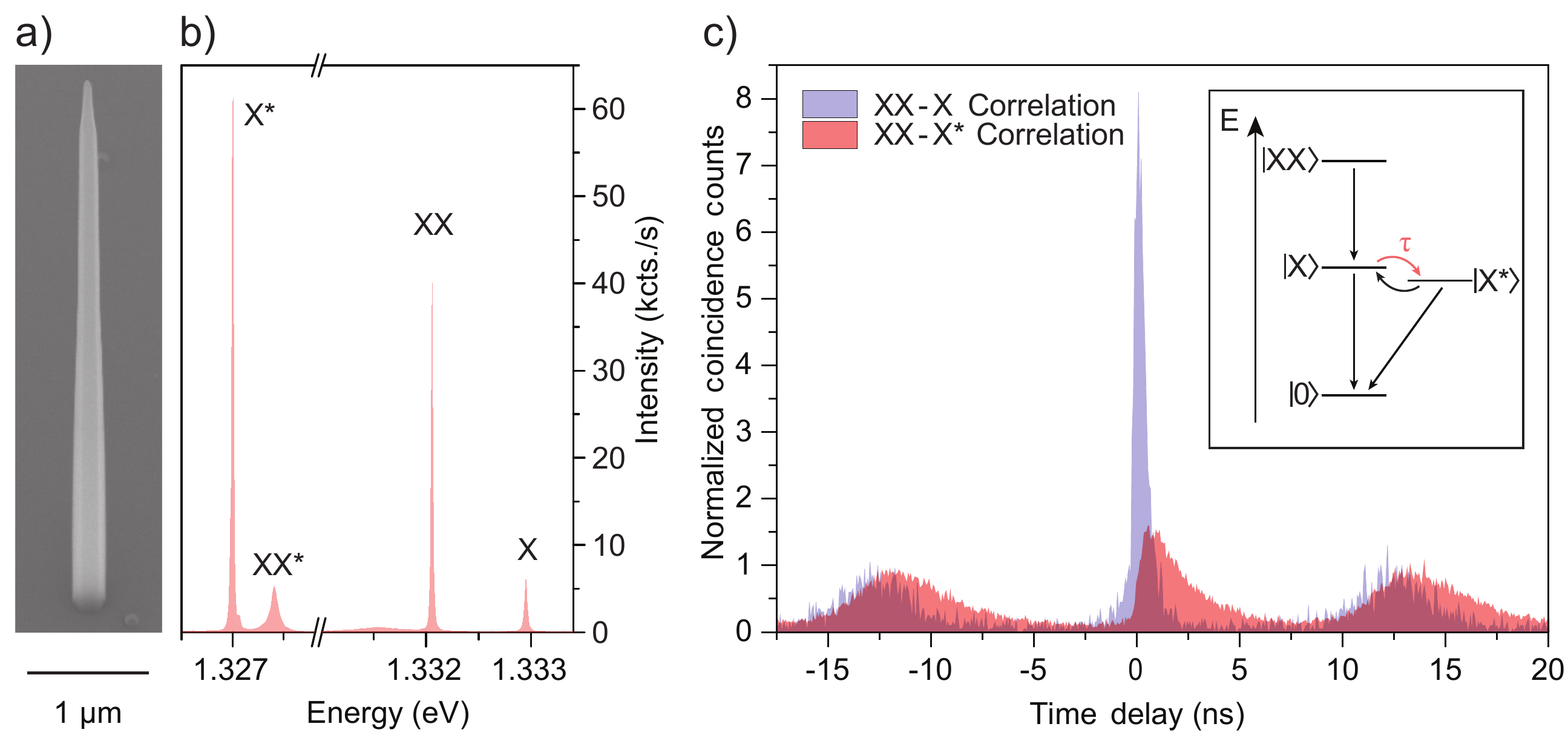}
\caption{\label{fig:fig1}a) SEM image of a pure wurtzite InP nanowire waveguide, containing a single InAsP quantum dot near the base. The tapering of the nanowire waveguide shell allows for efficient light extraction. b) Quantum dot spectrum of the s-shell transitions. c) Cross-correlation measurements of biexciton~(XX)--exciton~(X) cascade (blue) and biexciton--charged exciton~(X$^{*}$) cascade (red). The cross-correlation histograms are composed of 64\,ps time bins. A fast exciton decay with $T = 0.41\pm 0.08$\,ns starts after the biexciton trigger. The observed delay of the charged exciton emission after the biexciton trigger is attributed to the charging time $\tau = 0.39\pm 0.08$\,ns of the quantum dot. The inset schematically shows the charging effect and the two possible competing recombination pathways.}
\end{centering}
\end{figure*}

Here, we show that the emitted photon pairs from a quantum dot in a nanowire waveguide violate Bell's inequality without temporal post-selection. This violation is verified by the stringent Clauser-Horne-Shimony-Holt~(CHSH) measurement in the traditional basis~\cite{Clauser.Horne.ea:1969}. Importantly, by embedding our quantum dot source in a nanowire waveguide to steer the light emission, we collect two orders of magnitude more entangled photon pairs than typically reported from standard quantum dot structures.
In combination with the site-controlled growth in uniform arrays~\cite{Borgstrom.Immink.ea:2007}, the possibility to accurately transfer the nanowire sources into silicon-based photonic circuits with a nano-manipulator~\cite{Zadeh.Elshaari.ea:2016}, as well as the potential for electrical~\cite{Minot.Kelkensberg.ea:2007} and super-conducting contacting~\cite{Doh.Dam.ea:2005} of nanowires, our results on these nano-structured and bright entangled photon-pair sources with high fidelity highlight their great potential for future quantum emitter applications.


\section*{Photonic nanostructure}

We study InAsP quantum dots embedded in pure wurtzite InP nanowires that were grown by chemical beam epitaxy in a two-step growth process: vapor-liquid-solid and selective-area epitaxy. By using a self-alignment process, a gold particle is first centered in a circular opening of a SiO$_{\text{2}}$ mask on a (111)B substrate~\cite{Dalacu.Kam.ea:2009}. The gold particle is then used as a catalyst during the vapour-liquid-solid growth mode, defining the size and allowing for precise positioning of the nanowire core and quantum dot. Next, the growth conditions are modified to favor shell growth around the nanowire core via selective-area epitaxy in the circular opening of the SiO$_{\text{2}}$ mask. Further growing the nanowire results in a tapered waveguide shell around the nanowire core containing the quantum dot (see SEM image in Fig.~\ref{fig:fig1}\,a)). In such photonic structures, the photons emitted by the quantum dot are guided by the high refractive index material of the nanowire and a small tapering angle towards the nanowire tip results in efficient light extraction~\cite{Claudon.Bleuse.ea:2010}. In contrast to cavities, where achieving the necessary broadband frequency operation to realize bright entangled photon sources is extremely challenging, nanowire waveguides intrinsically provide this broadband frequency operation and have recently emerged as nanoscale sources with an excellent Gaussian far field emission profile~\cite{Munsch.Erratum.ea:2013} for near-unity fiber coupling~\cite{Bulgarini.Reimer.ea:2014}.
For the investigated nanowire we found a light extraction efficiency of $18 \pm 3$\,\%~\cite{Versteegh.Reimer.ea:2014}. This efficiency can be further increased by adding a mirror below the nanowire~\cite{Bleuse.Claudon.ea:2011, Reimer.Bulgarini.ea:2012}. A detailed description of the full fabrication process can be found in Ref.~\onlinecite{Dalacu.Mnaymneh.ea:2012}.

Fig.~\ref{fig:fig1}\,b) shows an s-shell emission spectrum of the investigated nanowire quantum dot. By performing cross-correlation measurements between the different transition lines, we found two competing recombination pathways in the quantum dot s-shell (illustrated in the inset of Fig.~\ref{fig:fig1}\,c)): the biexciton~(XX)--neutral exciton~(X) cascade~\cite{Benson.Santori.ea:2000,Moreau.Robert.ea:2001a} (blue curve in Fig.~\ref{fig:fig1}\,c)), and the biexciton~(XX)--charged exciton~(X$^{*}$) cascade~\cite{Shirane.Igarashi.ea:2010,Poem.Kodriano.ea:2010} (red curve in Fig.~\ref{fig:fig1}\, c)). Both measurements are taken under the same excitation conditions using the biexciton photon as the trigger at zero time delay. In the former case, the neutral exciton decays directly after the biexciton emission with a characteristic lifetime, $T$, of $0.41\,\pm\,0.08$\,ns. In contrast, the charged exciton decay is delayed by the charging time, $\tau$, of $0.39\,\pm\,0.08$\,ns, which is the time it takes for the quantum dot to capture another charge carrier. This competing recombination pathway explains the short decay time observed for the neutral exciton. Since the observed charging time is much faster than the period of the exciton spin precession ($\sim$3.5\,ns)~\cite{Versteegh.Reimer.ea:2014}, induced by the small fine-structure splitting~\cite{Stevenson.Hudson.ea:2008}, the effect of the spin precession on the two-photon correlations is small. Moreover, the short exciton lifetime filters out additional cross-dephasing events. Therefore, our short exciton lifetime eliminates two of the main reasons for applying temporal post-selection to increase the degree of entanglement~\cite{Ward.Dean.ea:2014,Huber.Predojevic.ea:2014}.

\section*{Compensation of birefringence}

\begin{figure*}[t]
\begin{centering}
\includegraphics[width=0.9\linewidth]{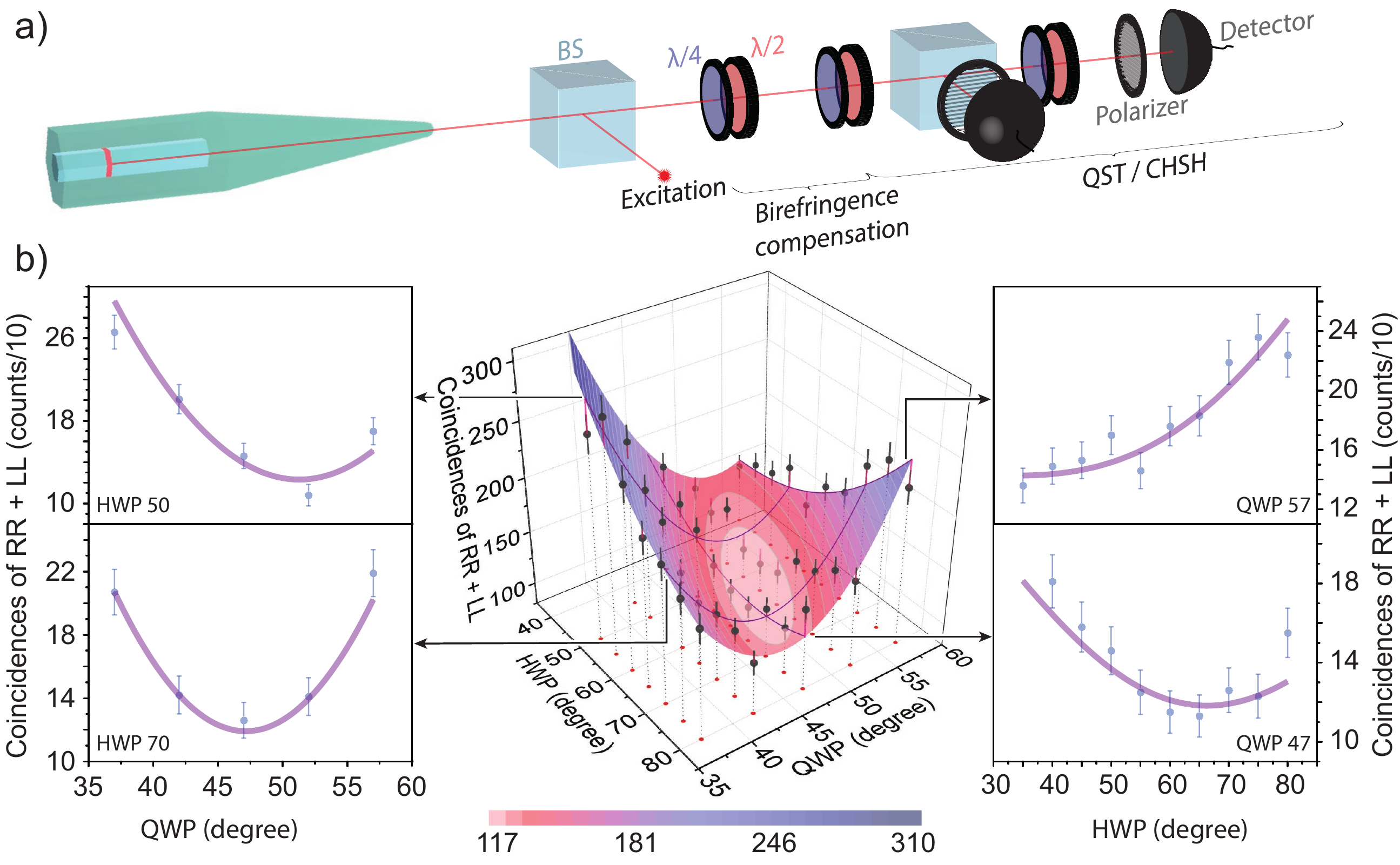}
\caption{\label{fig:fig2}a) Experimental setup for the Bell test and quantum state tomography (QST), including beam splitter (BS), half waveplates ($\lambda /2$), quarter waveplates ($\lambda /4$), polarizers and single-photon detectors. The first set of quarter and half waveplates is used to compensate the quantum state rotation introduced by the possible nanowire birefringence. b) 3D map showing the sum of coincidences at zero time delay for $|\text{RR}\rangle$ and $|\text{LL}\rangle$ correlations (black dots) as a function of quarter waveplate~(QWP) and half waveplate~(HWP) angles. Using a color-coded cosine surface fit we found the optimum waveplate configuration to compensate for the state rotation (minimum correlations in $|\text{RR}\rangle$ and $|\text{LL}\rangle$). Slices of the 3D map at constant HWP (QWP) angles are shown on the left (right) illustrating the good agreement between the fit and data.}
\end{centering}
\end{figure*}

Typically the entangled state obtained for conventional quantum dots is $(|\text{HH}\rangle + |\text{VV}\rangle)/\sqrt{2}$~\cite{Akopian.Lindner.ea:2006,Young.Stevenson.ea:2006,Hafenbrak.Ulrich.ea:2007,Juska.Dimastrodonato.ea:2013}. However, as previously reported~\cite{Versteegh.Reimer.ea:2014}, we observed a different entangled state from the quantum dot embedded in the waveguiding shell.
This more general entangled state was described as $(|\text{JJ}\rangle + |\text{WW}\rangle)/\sqrt{2}$, where J and W are arbitrary orthogonal vectors in the Poincare sphere, describing an elliptical polarization state.
This rotation of the conventional state is most likely introduced by birefringence of the nanowire waveguide arising from an asymmetric shape. Using a set of waveplates we can compensate for this rotation and bring the entangled biexciton-exciton quantum state back to $(|\text{HH}\rangle + |\text{VV}\rangle)/\sqrt{2}$, allowing us to perform the CHSH measurements in the traditional basis~\cite{Clauser.Horne.ea:1969}.

In the following we optimize the waveplates to compensate for the observed state rotation. A schematic of the experimental setup is depicted in Fig.~\ref{fig:fig2}\,a) where the emitted photons from the nanowire quantum dot are first sent through a $\lambda /4$~(QWP) and a $\lambda /2$~(HWP) waveplate before entering the analyzing polarization dependent cross-correlation setup (either for quantum state tomography~(QST) or the Bell test~(CHSH)). If the quantum state of the photon pair is of the form $(|\text{JJ}\rangle + |\text{WW}\rangle)/\sqrt{2}$, with J and W being orthogonal polarization vectors, any state rotation can be compensated by minimizing the sum of the measured coincidences in $|\text{RR}\rangle$ and $|\text{LL}\rangle$ using these waveplates. Fig.~\ref{fig:fig2}\,b) shows a three-dimensional~(3D) map of these coincidences as a function of the QWP and HWP angles. The data is shown as black dots and the 3D cosine function fit of the data is color-coded. A clear minimum is observed in the light pink region, indicating the optimum waveplate angles required to compensate for the state rotation. The excellent result of the fit is emphasized in four cuts (dark purple) through the 3D map, displayed next to the map.
The extracted QWP and HWP angles at the minimum of the surface fit will rotate the quantum state back to $(|\text{HH}\rangle + |\text{VV}\rangle)/\sqrt{2}$. This can be either verified by a full quantum state tomography or a simple fidelity approximation to the $(|\text{HH}\rangle + |\text{VV}\rangle)/\sqrt{2}$ state as shown in section: Improved Bell's inequality violation.

\section*{CHSH violation}

After our successful compensation of the birefringence, we can perform the traditional four-parameter CHSH test, the standard test of local realism and device-independent secure quantum communication, to violate Bell's inequality.

\begin{figure*}[t]
\begin{centering}
\includegraphics[width=0.9\linewidth]{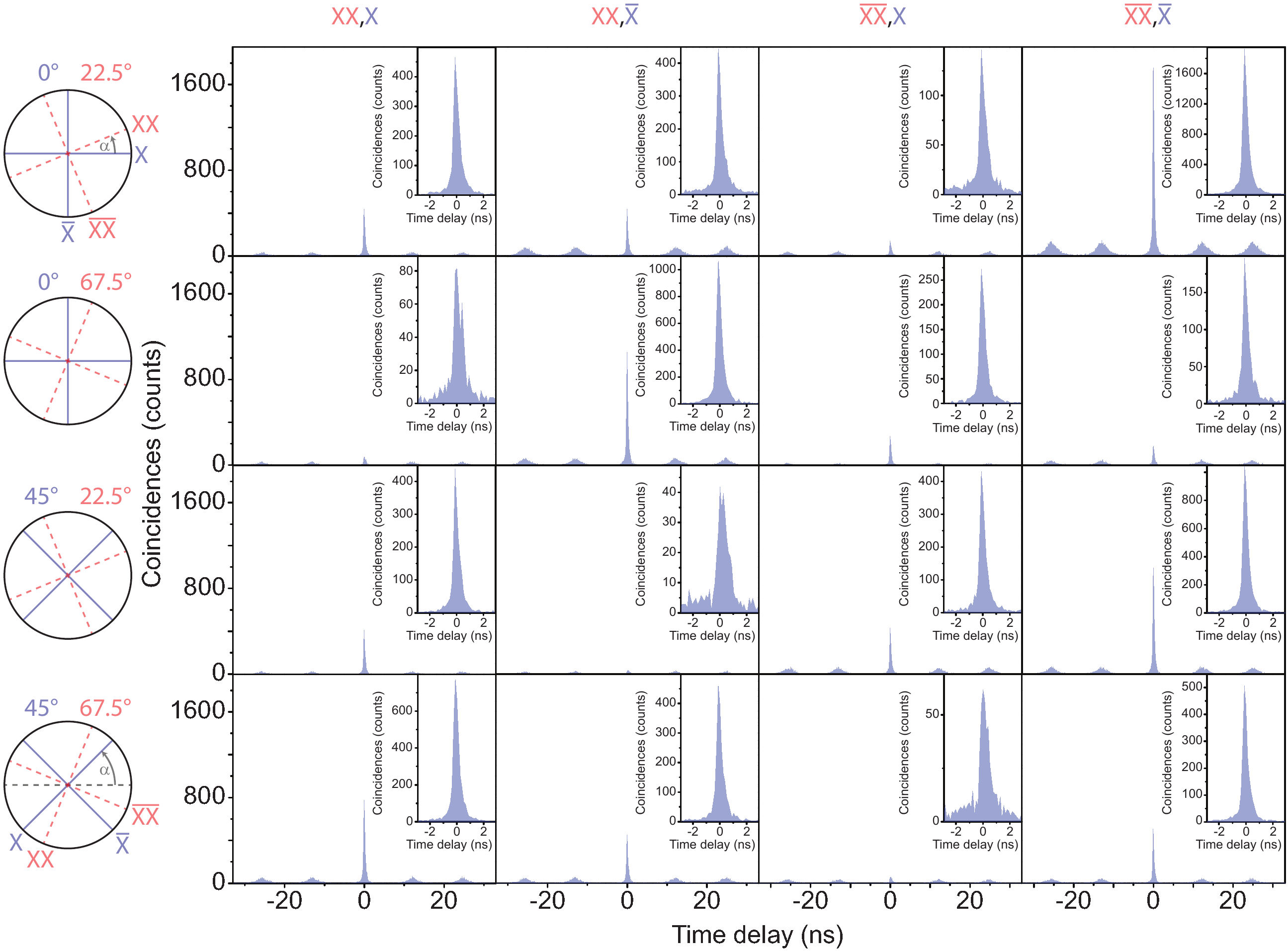}
\caption{\label{fig:fig3}The sixteen cross-correlation measurements needed for the violation of the Clauser-Horne-Shimony-Holt inequality. The angles on the left specify how much the polarization detection angle $\alpha$ for the exciton (blue) and biexciton (red) is rotated with respect to the rectilinear basis in the laboratory reference frame. For each set of rotation the degree of correlation $C_{b}$ is extracted by measuring all four combinations between the biexciton ($\text{XX}$ and $\overline{\text{XX}}$) and exciton ($\text{X}$ and $\overline{\text{X}}$) photons. The insets show a magnification of the zero time delay peak.}
\end{centering}
\end{figure*}

The CHSH test requires 16 non-orthogonal cross-correlation measurements taken under the same experimental conditions. In the CHSH test the biexciton and exciton detection bases are independently rotated with respect to the rectilinear laboratory reference frame. The rotation angles $\alpha$ are given in real space angles in the schematic on the left of Fig.~\ref{fig:fig3}. The values $22.5^{\circ}$, 67.5$^{\circ}$ in red (0$^{\circ}$, 45$^{\circ}$ in blue) correspond to the polarization angles added to the biexciton (exciton) polarization detection bases. To extract the degree of correlation $C_{b}$ in all four possible combinations of these bases, we measure the coincidences between the biexciton (XX and $\overline{\text{XX}}$) and exciton (X and $\overline{\text{X}}$) photons. The degree of correlation $C_{b}$ in one polarization basis $b$ is extracted from the polarization dependent cross-correlation measurements (shown in Fig.~\ref{fig:fig3}): 
\begin{equation}
\label{equ:C}
C_{b}=\frac{g_{xx,x}+g_{\overline{xx},\overline{x}}-g_{xx,\overline{x}}-g_{\overline{xx},x}}{g_{xx,x}+g_{\overline{xx},\overline{x}}+g_{xx,\overline{x}}+g_{\overline{xx},x}},
\end{equation}

\noindent 
where $g_{xx,x}$ and $g_{\overline{xx},\overline{x}}$ ($g_{xx,\overline{x}}$ and $g_{\overline{xx},x}$) are the coincidences of the co-polarized (cross-polarized) biexciton and exciton photons.
Each measurement depicted in Fig.~\ref{fig:fig3} was integrated for 7200\,s and the histograms are composed of 128\,ps time bins. The insets magnify the central peak at zero time delay. We calculate the Bell parameter with:
\begin{equation}
\label{equ:S}
S_{\text{CHSH}} = C_{22.5^{\circ}, 0^{\circ}}-C_{67.5^{\circ}, 0^{\circ}}+C_{67.5^{\circ}, 45^{\circ}}+C_{22.5^{\circ}, 45^{\circ}} \leq 2.
\end{equation}

\begin{table}
\centering 
\caption{\label{table:table1}Extracted Bell parameter from the CHSH measurements in the linear plane of the Poincar\'{e} sphere. The percentage of the correlation events taken into account for a certain time window is given in the second column.}
    \begin{tabular}{ c  c  c }
    \hline
     Time window (ns) & Counts (\%) & S$_{\text{CHSH}}$ \\ 
		\hline 
  4.48 & 100 & $2.07  \pm 0.02$ \\
	1.41 & 81 & $2.17  \pm 0.02$ \\
	0.38 & 38 & $2.28  \pm 0.03$ \\
	0.13 & 12 & $2.35  \pm 0.06$ \\
   \hline
    \end{tabular}
\end{table}

In Table~\ref{table:table1} we present the calculated Bell parameters extracted from the CHSH measurements in the linear plane of the Poincar\'{e} sphere for four different time windows.
Importantly, taking 100\,\% of the emitted photons into account, the calculated Bell parameter $S_{\text{CHSH}}$ extracted from the CHSH measurements is $2.07 \pm 0.02$. To violate the traditional CHSH inequality with the necessary certainty needed for applications in quantum key distribution we have to employ temporal post-selection. As shown in Table~\ref{table:table1} we already achieve a sufficient violation by 8.5 standard deviations in a time window which is still 3.4 times longer than the excitonic lifetime, discarding only 19\,\% of the emitted photon pairs. However, temporal post-selection complicates the implementation of quantum cryptography schemes severely. Therefore, we explore a quasi-resonant excitation scheme to significantly violate Bell's inequality without the need for temporal post-selection.

\section*{Quasi-resonant excitation}

As shown in Table~\ref{table:table1} temporal post-selection still increases the S$_{\text{CHSH}}$ parameter even though the fine structure induced exciton dephasing at short time windows is negligible. Therefore, the false correlations in the CHSH measurements most likely originate from a re-excitation of the quantum dot and not from dephasing events. 

\begin{figure}[ht]
\begin{centering}
\includegraphics[width=0.49\linewidth]{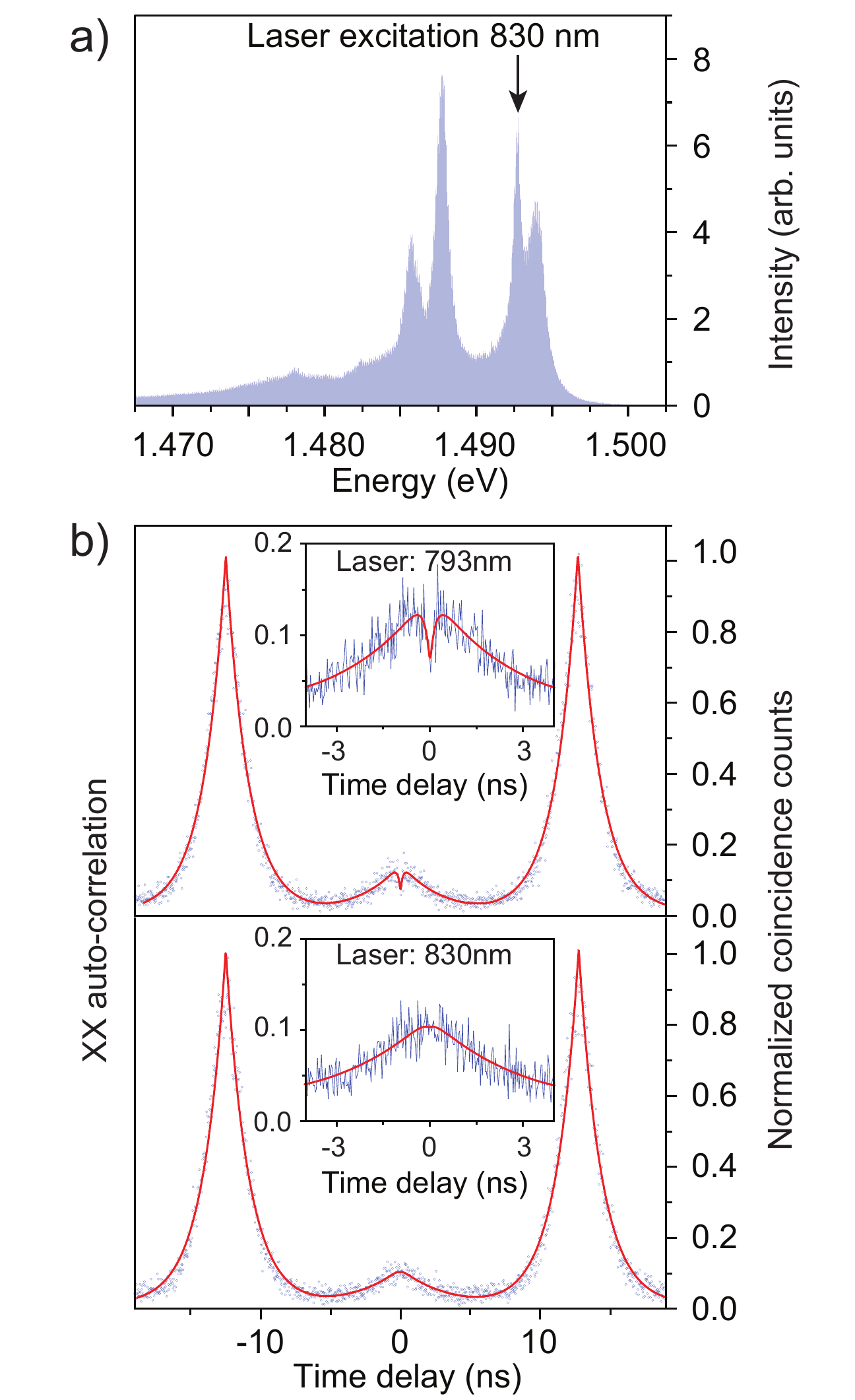}
\caption{\label{fig:fig4}a) Photoluminescence spectrum of the wurtzite InP nanowire resonances. The sharp resonance at 830\,nm was addressed with a ps-pulse laser to excite the quantum dot quasi-resonantly. 
b) Auto-correlation histograms (time bins 32\,ps) of the biexciton for above-band excitation at 793\,nm (top) and quasi-resonant excitation in the InP nanowire wurtzite resonance at 830\,nm (bottom). As shown in the insets the volcano-shaped dip at zero time delay, characteristic for re-excitation processes, is significantly reduced for the quasi-resonant excitation scheme in the lower panel.}
\end{centering}
\end{figure}

Re-excitation can be reduced by minimizing the density of free charge carries in the nanowire, either by reducing the excitation power or by a more resonant excitation scheme. 
Fig.~\ref{fig:fig4}\,a) shows the emission spectrum of the wurtzite InP nanowire resonances suitable for quasi-resonant excitation of the quantum dot. 
To verify our assumption that the false coincidences in the CHSH measurement originate from re-excitation of the quantum dot within one excitation laser pulse, we perform auto-correlation measurements on the biexciton. Fig.~\ref{fig:fig4}\,b) depicts two of these measurements at different excitation laser wavelengths, while keeping the measured photon flux on the detectors constant. To maintain the same photon flux the laser power was reduced by 30\,\% for the quasi-resonant excitation. The top graph of Fig.~\ref{fig:fig4}\,b) shows the result for above-band excitation at a wavelength of 793\,nm. The non-zero anti-bunching peak at zero time delay partly originates from re-excitation of the biexciton during the same laser excitation pulse. The signature of re-excitation is the volcano-shaped peak shown in the inset, leading to increased correlations near zero time delay. The dip in the zero time delay peak is caused by the biexciton needing time to decay before the re-excitation process. These additional biexciton photons start an additional cascade, which is independent of the first one, leading to false correlation counts in the CHSH measurement and to a reduction of the measured degree of entanglement. Excitation of the wurtzite nanowire resonance at 830\,nm (bottom graph of Fig.~\ref{fig:fig4}\,b)) significantly reduces the amount of re-excitation, indicated by the missing volcano-shape peak at zero time delay, as shown in the inset.
The quasi-resonant excitation results in an overall improvement of the g$^{(2)}$(0) by approximately 15\,\%. The remaining coincidences at zero time delay can be explained by background emission.

\section*{Improved Bell's inequality violation}
\label{sec:bell}

\begin{figure}[ht]
\begin{centering}
\includegraphics[width=0.49\linewidth]{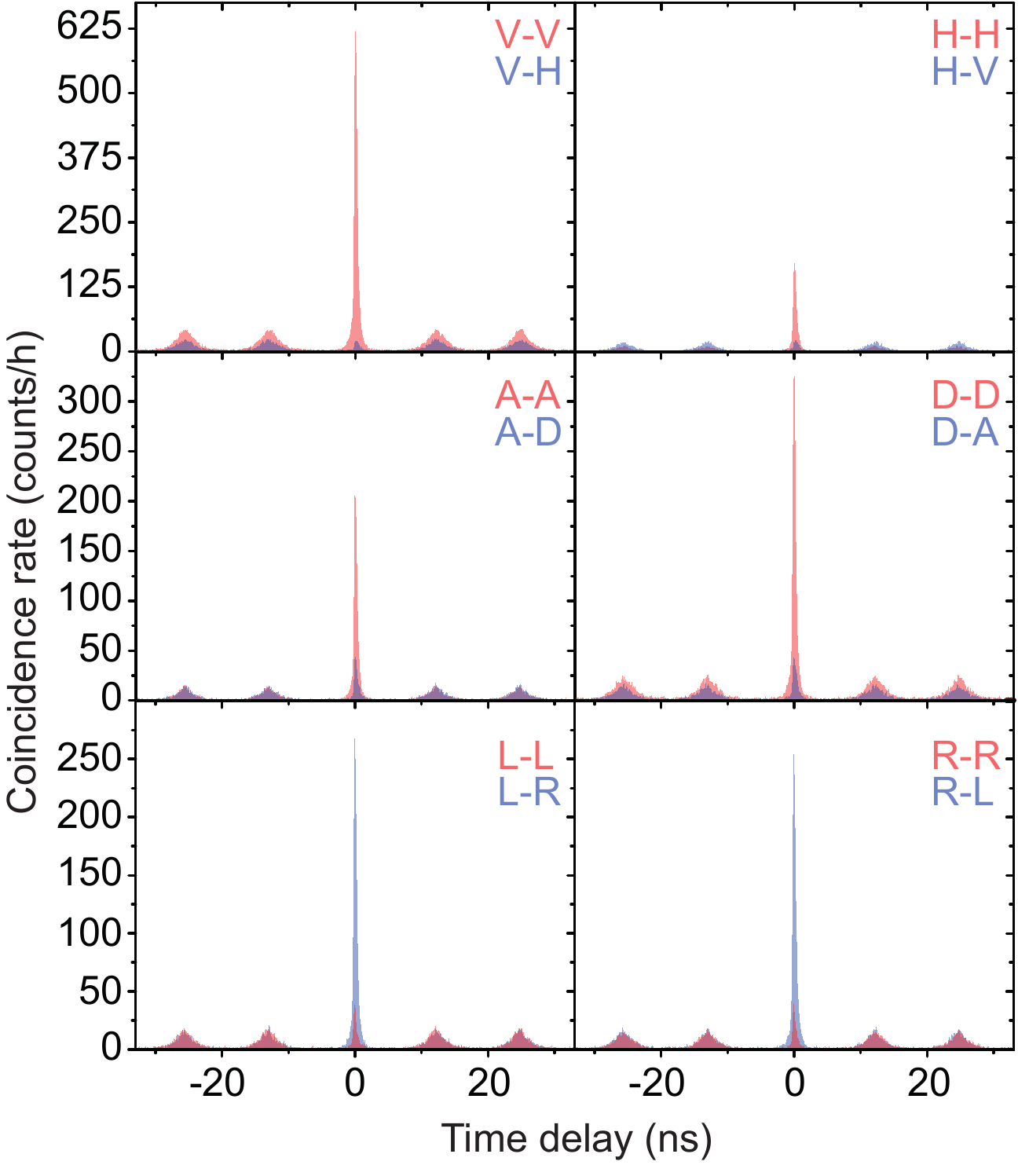}
\caption{\label{fig:fig5}Twelve cross-correlation measurements between the biexciton and exciton for different polarization detection bases, from top to bottom: rectilinear, diagonal and circular basis. The first (second) letter in the graphs stands for the polarization detection angle of the biexciton (exciton). The fidelity to the $(|\text{HH}\rangle + |\text{VV}\rangle)/\sqrt{2}$ state is $F = 0.817 \pm 0.002$ without temporal post-selection. These measurements are used to extract the Bell parameters in three orthogonal planes of the Poincar\'{e} sphere, given in Table~\ref{table:table2}.}
\end{centering}
\end{figure}

Using the quasi-resonant excitation scheme we characterize the enhanced entangled photon-pair emission of our photonic nanostructure. 
Fig.~\ref{fig:fig5} shows the twelve non-normalized polarization dependent cross-correlation measurements~\cite{Shields.Stevenson.ea:2009} in three different bases (rectilinear, diagonal, and circular) needed for the fidelity approximation. The histograms are composed of 128\,ps time bins and the first (second) letter denotes the biexciton (exciton) polarization. Strong positive correlations are visible in the rectilinear and diagonal basis and negative correlations in the circular basis, indicative of entanglement. 
From these measurements we calculate the fidelity, $F$, to the state $(|\text{HH}\rangle + |\text{VV}\rangle)/\sqrt{2}$~\cite{Hudson.Stevenson.ea:2007}:
\begin{equation}
\label{equ:F}
F = (1+C_{\text{rectilinear}}+C_{\text{diagonal}}-C_{\text{circular}})/4.
\end{equation}
We obtain a fidelity of $F=0.850\,\pm\,0.009$ to the state $(|\text{HH}\rangle + |\text{VV}\rangle)/\sqrt{2}$ for a time window of 0.13\,ns. 
In Table~\ref{table:table2} the fidelity to the state $(|\text{HH}\rangle + |\text{VV}\rangle)/\sqrt{2}$ for four different time windows, including the percentage of the correlation events taken into account, is given. For longer time windows the importance of our elegant quasi-resonant excitation of the nanowire is apparent. The fidelity is still as high as $F=0.817\,\pm\,0.002$ for the full time window, a significant improvement as compared to above-band excitation, where $F=0.762\,\pm\,0.002$. In addition to the reduced re-excitation induced coincidences, which occur at non-zero time delay, the quasi-resonant excitation should also reduce the amount of charge carriers surrounding the quantum dot. These reduced charge carriers in the environment potentially lead to fewer spin scattering and cross-dephasing events. 

\begin{table*}[ht]
\centering
\caption{\label{table:table2}Calculated fidelity to $(|\text{HH}\rangle + |\text{VV}\rangle)/\sqrt{2}$ for four different time windows. The percentage of the correlation events taken into account for a certain time window is given in the second column. The calculated Bell parameters for all three bases combinations are always above the classical limit of 2.}
    \begin{tabular}{ c  c  c  c  c  c }
    \hline
     Time window (ns) & Counts (\%) & Fidelity to $(|\text{HH}\rangle + |\text{VV}\rangle)/\sqrt{2}$ & S$_{rc}$ & S$_{dc}$ & S$_{rd}$\\ 
		\hline 
  4.48 & 100 & $0.817 \pm 0.002$  & $2.25 \pm 0.01$ & $2.01 \pm 0.01$ & $2.16 \pm 0.01$\\
	1.41 & 88 & $0.829 \pm 0.002$ & $2.26 \pm 0.01$ & $2.04 \pm 0.01$ & $2.21 \pm 0.01$\\
	0.38 & 50 & $0.847 \pm 0.003$ & $2.36 \pm 0.01$ & $2.10 \pm 0.02$ & $2.30 \pm 0.01$\\
	0.13 & 18 & $0.850 \pm 0.005$ & $2.37 \pm 0.02$ & $2.10 \pm 0.03$ & $2.32 \pm 0.02$\\
   \hline
    \end{tabular}
\end{table*}

We re-emphasize that the high degree of polarization-entangled photon pairs from our nanowire quantum dot, for the full time window, is benefited by the charging time $\tau$. Controlling this effect could be achieved in future work with a gate to switch the photon emitting state population. The optimum gating time would depend on the excitonic fine-structure splitting. A similar gating technique was used to increase the visibility of two-photon interference~\cite{Bennett.Patel.ea:2008}.

Using the degree of correlation $C_{b}$ extracted from the four corresponding cross-correlation measurements in that basis $b$, we can calculate the Bell parameters for our source. However, compared to the traditional CHSH measurement, no test of local hidden-variable theories is made, simply accepting the non-locality of quantum mechanics~\cite{2008arXiv0801.4549T}. Following the work of Young and co-workers~\cite{Young.Stevenson.ea:2009} three different Bell parameters: $S_{\text{RC}}$ (rectilinear-circular), $S_{\text{DC}}$ (diagonal-circular), and $S_{\text{RD}}$ (rectilinear-diagonal), are given in Table~\ref{table:table2}. Each one corresponds to a measurement in one of the three orthogonal planes of the Poincar\'{e} sphere. A value above 2 in these Bell parameters, as shown for all time windows for our entangled photon-pair source, confirms the non-classical character of the two-photon state. Employing this novel quasi-resonant excitation scheme we achieve a violation of Bells inequality in all three bases without temporal post-selection. In the rectilinear-circular basis  a violation of Bell's inequality by 25 standard deviations is achieved. We violate by 14 standard deviations if averaged over all bases. In contrast to previous works that violate Bell's inequality with optically active self-assembled quantum dots~\cite{Young.Stevenson.ea:2009,Kuroda.Mano.ea:2013,Trotta.Wildmann.ea:2014}, our nanowire quantum dot photon emission is slightly polarized. This effect has to be taken into account by performing four instead of two cross-correlation measurements to extract the degree of correlation in each basis, $b$. This more general approach to determine the Bell parameters is then not limited to unpolarized quantum light sources.

\begin{figure}[th]
\begin{centering}
\includegraphics[width=0.49\linewidth]{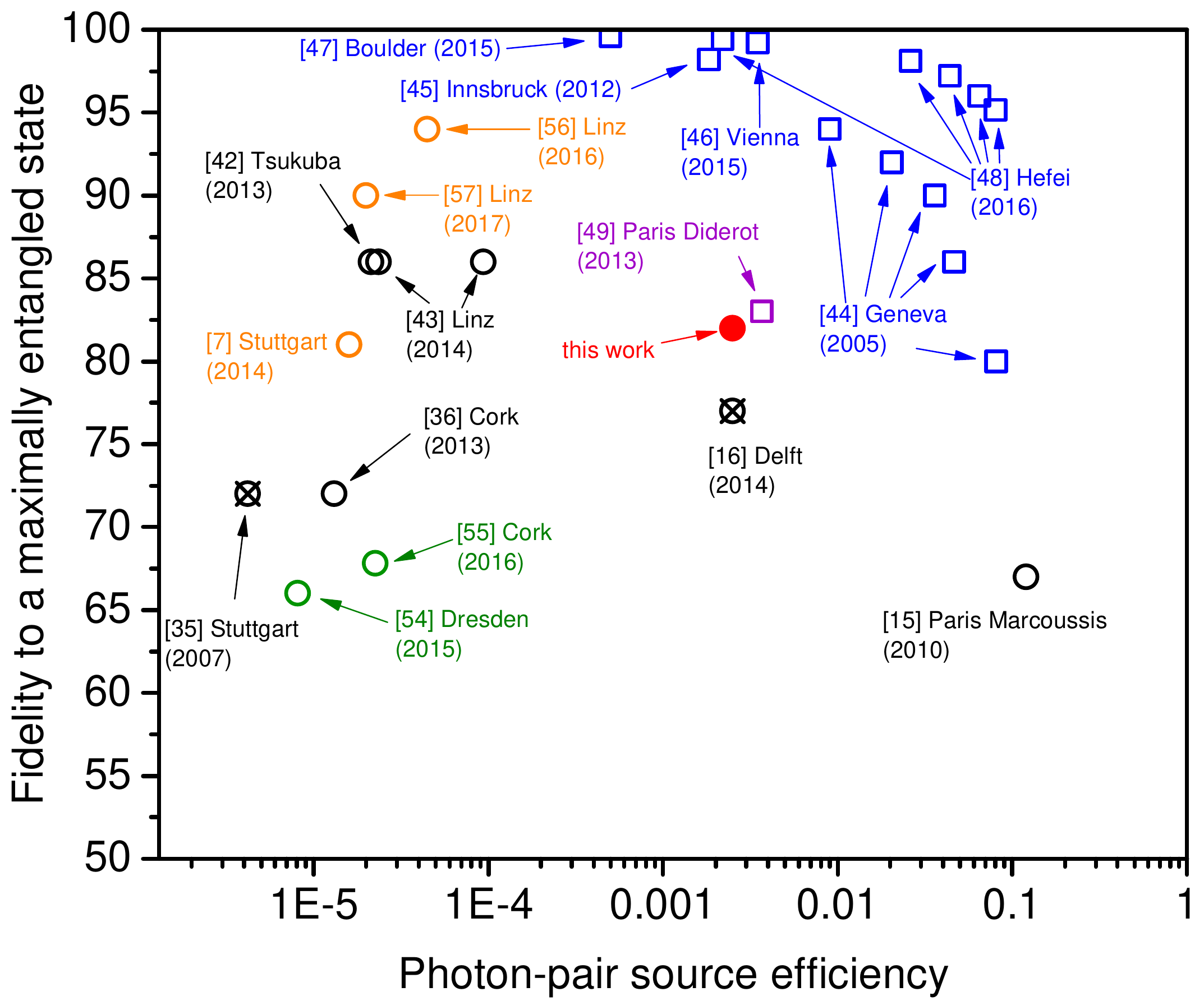}
\caption{\label{fig:fig6} Comparison of the entanglement fidelity and photon-pair source efficiency of different parametric down-conversion (squares) and quantum dot-based (circles) pulsed entangled photon-pairs sources. The photon-pair source efficiency is defined as the probability per excitation pulse to collect a photon-pair at the first lens or fiber. Note that, the fidelity of the quantum dot-based sources is calculated for the non post-selected case, where all emitted photons are taken into account. For the quantum dot sources, the black color corresponds to above-band excitation, the green color to electrical excitation, the orange color represents resonant biexciton excitation, and the red color indicates quasi-resonant excitation. The crossed black circles symbolizes above-band excited sources which have also been measured under quasi-/resonant excitation conditions. 
References of the depicted data points are given in brackets next to the symbols.}
\end{centering}
\end{figure}

In Fig.~\ref{fig:fig6}, we compare the fidelity to a maximally entangled state as a function of the photon-pair source efficiency for quantum dot and parametric down-conversion sources under pulsed excitation. Here, the photon-pair source efficiency is defined as the probability of collecting a photon-pair per excitation pulse into the first lens or fiber. We note that the photon-pair source efficiency includes all losses up to the first collection optics (e.g. extraction efficiency, internal quantum efficiency, excitation probability, etc...). Parametric down-conversion sources are represented in blue squares~\cite{Scarani.DeRiedmatten.ea:2005,Predojevic.Grabher.ea:2012,Giustina.Versteegh.ea:2015,Shalm.Meyer-Scott.ea:2015,Wang.Chen.ea:2016}, while the quantum dot sources are represented as circles in black (above-band excitation), green (electrical excitation), orange (resonant excitation) and red (quasi-resonant excitation). A detailed explanation of how each individual data point was derived, is given in the methods. For parametric down-conversion sources, the entanglement fidelity has a mutual dependence on the photon-pair source efficiency due to the probabilistic nature of the pair generation efficiency. Therefore, these sources cannot have near-unity photon-pair source efficiency while maintaining a high fidelity. As a result the entanglement fidelity for higher photon-pair collection efficiency rapidly degrades. In contrast, since quantum dot sources are deterministic and can emit photon-pairs on demand, they have the potential to reach near-unity source efficiency without negatively influencing the entanglement fidelity. The data presented in our work shows the highest combination of photon pair collection efficiency and fidelity for a quantum-dot based source (red circle) and is comparable to state-of-the-art probabilistic AlGaAs semiconductor waveguide sources (purple square)~\cite{Orieux.Eckstein.ea:2013}. Still, the entanglement fidelity is lower than for parametric down-conversion sources for similar source efficiencies. However, using two-photon resonant excitation we strongly believe that both the entangled photon source fidelity (by minimizing multi-photon emission) and efficiency (due to the near-unity biexciton preparation and the suppression of the trion transition) could be boosted further towards unity. This influence of two-photon resonant excitation can be identified by comparing both studies from Stuttgart (orange circle~\cite{Muller.Bounouar.ea:2014} and crossed black circle \cite{Hafenbrak.Ulrich.ea:2007}), for example, since the same source was used.

Due to the high brightness of our nanowire quantum dot (200k entangled photon-pairs per second collected into the first lens, corresponding to a $\sim$10\,Hz two-photon coincidence detection rate), we want to highlight its potential as a referee source~\cite{Nawareg.Muhammad.ea:2015} for measurement device-independent cryptography and quantum dot-based relay schemes~\cite{Varnava.Stevenson.ea:2016}. Assuming a setup consisting of four state-of-the-art detectors with a quantum efficiency of 70\%, a dichroic mirror for the confocal microscopy, and an efficient way to separate the biexciton and exciton our source would violate Bell's inequality in the rectilinear-circular basis in less than 1 second by 5 standard deviations, which is considered a significant test on the security of the communication channel. Our source can complete this task so fast due to the combination of the strong degree of entanglement and high photon-pair source efficiency. However, such applications also require a high degree of photon indistinguishability. Up to now, only post-selected two-photon interference measurements between consecutively emitted photons from nanowire quantum dots showed high visibilities~\cite{Chen.Zadeh.ea:2016,Reimer.Bulgarini.ea:2016}. The visibilities were limited in these works by timing jitter and spectral diffusion under the above-band excitation process. We expect that resonant excitation will enhance the two-photon interference visibility similarly to self-assembled quantum dots,  bringing nanowire quantum dots one step closer to applications.

\section*{Conclusion}
We have demonstrated a bright entangled photon source in tapered nanowires waveguides that violates Bell's inequality by 25 standard deviations without the need for temporal post-selection. This is achieved by quasi-resonant excitation in the wurtzite nanowire, which reduces re-excitation induced multi-photon emission of the biexciton state, while maintaining the high entangled-photon pair flux. We compensated for the quantum state rotation possibly introduced by the nanowire birefringence, reconstructing the state $(|\text{HH}\rangle + |\text{VV}\rangle)/\sqrt{2}$ with a fidelity of $F=0.817 \pm 0.002$ without temporal post-selection. Studying the correlations between competing cascaded recombination channels revealed an intrinsic charging of the quantum dot on time scales shorter than the fine structure splitting induced excitonic spin precession time, explaining the high degree of entanglement observed. Our work highlights the applicability of novel nanowire quantum dots as bright entangled photon-pair sources suitable for measurement device-independent quantum cryptography and quantum information processing.

\section*{Methods}
\label{sec:methods}
\subsection*{Comparison of different entangled photon-pair sources}
\label{sec:comparison}

To make a comparison between different types of polarization-entangled photon-pair sources a common basis has to be found. We chose to compare the fidelity to a maximally entangled state as a function of the photon-pair source efficiency. For the quantum dot related publications the efficiency is given as the average number of photon pairs per excitation pulse collected into the first lens, whereas for the parametric down-conversion sources it is the same figure of merit but collected into the first fiber. The photon-pair source efficiency includes the collection efficiency and the pair generation efficiency. In quantum dot studies the source efficiency is typically calculated from the excitation pulse rate, the setup efficiency and the count rates on the detectors. It directly includes the internal pair generation efficiency of the investigated quantum dot. The pair generation efficiency varies from dot to dot, due to different recombination channels, and depends on the excitation conditions. For parametric down-conversion sources the internal pair generation efficiency depends on the excitation power. Due to the probabilistic nature of the generation, higher excitation power also leads to multi-pair generation reducing the source fidelity. In Fig.~\ref{fig:fig6} we only compare fidelities from pulsed sources and the quantum dot based data were all obtained without additional temporal post-selection. 
For each cited work an explanation of how the data points for Fig.~\ref{fig:fig6} were obtained is given below:


Geneva [\onlinecite{Scarani.DeRiedmatten.ea:2005}]: All the required data was given in the manuscript. We used the data where both photons were spectrally filtered. The Fidelity measured to a maximally entangled state was measured in the timebin basis. 

Innsbruck [\onlinecite{Predojevic.Grabher.ea:2012}]: The fidelity and the measured photon-pair rate are given in the manuscript. The data from the manuscript was corrected with the setup efficiency of 0.405 to obtain the photon-pair source efficiency into the first lens.

Vienna [\onlinecite{Giustina.Versteegh.ea:2015}]: The measured visibilities in the linear and diagonal bases are 0.997 and 0.993, respectively. We estimated the fidelity (0.992) based on the two given visibilities. 

Boulder [\onlinecite{Shalm.Meyer-Scott.ea:2015}]: The visibilities in the linear and diagonal bases are 0.999 and 0.996. We estimated the fidelity (0.996) based on the two given visibilities.  

Hefei [\onlinecite{Wang.Chen.ea:2016}]: All the required data was given in the manuscript and its supplemental information. We note that the given overall efficiency for a single photon (0.42) is corrected for collection efficiency which is not the case for all other data points. Therefore, we back calculated the efficiency not including the collection loss correction: $0.42 / 0.7 = 0.6$, to make a fair comparison.

Paris Diderot [\onlinecite{Orieux.Eckstein.ea:2013}]: In the paper the visibility (0.83) and source efficiency (0.007) was given. However, the efficiency is corrected for facet losses at the end of the waveguide. This is not the case for all the other data points in the graph. Therefore, we back calculated the efficiency not including the facet loss correction. The real number for the photon-pair source efficiency of this source is $0.007 \times 0.5329 = 0.0037$. 

Paris Marcoussis [\onlinecite{Dousse.Suffczynski.ea:2010}]: All the required data was given in the manuscript. We used the data without any kind of time gating.

Delft [\onlinecite{Versteegh.Reimer.ea:2014}]: All required data was given in the manuscript (crossed black circle in Fig.~\ref{fig:fig6}). The same source was used in our current work (red circle).


Linz [\onlinecite{Trotta.Wildmann.ea:2014}]: The photon-pair source efficiency is estimated based on the setup efficiency, resulting in a collection efficiency into the first lens of 1\,\%. This estimation is compared with the measured count rates on the detectors and the used equipment. The estimation is double checked with the theoretical collection efficiency of the used sample structure (4\,\%). Based on the collection efficiency into the first lens and the biexciton and exciton detector counts, the photon-pair source efficiency is calculated. The difference in photon-pair source efficiency for the two plotted data points is based on the different biexciton and exciton detector counts for the two investigated quantum dots. It gives a good estimate of the spread in the internal pair generation efficiency for non-resonant excitation.

Tsukuba [\onlinecite{Kuroda.Mano.ea:2013}]: The photon-pair source efficiency is based on an estimate of the setup efficiency, resulting in a collection efficiency into the first lens of 0.8\,\%. This estimation is compared with the measured count rates on the detectors and the used equipment. The estimation is double checked with the theoretical collection efficiency of the used sample structure (0.81\,\%). Based on the collection efficiency into the first lens and the biexciton-exciton intensity ratio the photon-pair source efficiency is calculated.

Stuttgart [\onlinecite{Muller.Bounouar.ea:2014},\onlinecite{Hafenbrak.Ulrich.ea:2007}]: All required data was given in the manuscript reporting on two-photon excitation (orange circle in Fig.~\ref{fig:fig6}). For this resonant excitation the internal pair generation efficiency per excitation pulse was as high as $86\pm 8$\,\%. The same source and setup was used in their previous work. Based on the setup efficiency and the biexciton and exciton count rates, the photon-pair source efficiency is calculated for the above-band excitation result (crossed black circle).

Cork [\onlinecite{Juska.Dimastrodonato.ea:2013}]: The photon-pair source efficiency is based on an estimate of the setup efficiency (1\,\% from source to one detector), resulting in a collection efficiency into the first lens of 0.75\,\%. To verify this estimation, it is double checked with the measured count rates on the detectors and the used equipment. Based on the setup efficiency and the biexciton and exciton count rates the photon-pair source efficiency is calculated.

Dresden [\onlinecite{Zhang.Wildmann.ea:2015}]: The quantum dot is excited electrically with a repetition rate of 400MHz. The photon-pair source efficiency is calculated based on the measured setup efficiency of 1.1\,\% and the detector count rates. Even though the structure has a high extraction efficiency of approx. 5\,\% the electrical injection is not very efficient. Per electrical cycle approx. 0.3\,\% photon pairs are generated in the quantum dot (assuming 100\,\% internal quantum efficiency of the quantum dot).  For comparison Slater et al. \cite{Salter.Stevenson.ea:2010} reported a probability of 3\,\% to generate an entangled photon pair per electrical injection cycle, however with low extraction efficiency. If one could remove all the addition losses due to the low efficiency of the electrical excitation, the work of Dresden would have a photon pair source efficiency of approx. 0.025\,\%.

Cork [\onlinecite{Chung.Juska.ea:2016}]: All the required data (photon-pair detection rate, repetition rate of the excitation, setup efficiency for detecting a single-photon) was given in the manuscript. We used the data without any kind of time gating.

Linz [\onlinecite{2016arXiv161006889H}]: The photon-pair source efficiency is calculated from the measured setup efficiency of $\approx 0.007$, the detector count rates for X and XX ($\approx 3750$\,Hz each) and the repetition rate (80\,MHz) of the laser. 

Linz [\onlinecite{2017arXiv170107812R}]: The quantum dot is excited via phonon-assisted two-photon excitation with a repetition rate of 80\,MHz. The photon-pair source efficiency is calculated from the measured setup efficiency of $\approx 0.007$, the repetition rate, and the detector count rates for X and XX ($\approx 2500$\,Hz each). 

\subsection*{Error calculation}
The given errors for values based on correlation measurements are initially obtained from the square root of the recorded coincidences and propagated quadratically.

\section*{Funding Information}
This research was supported by the Dutch Foundation for Fundamental Research on Matter (FOM projectruimte 12PR2994), Industry Canada, ERC, and the European Union Seventh Framework Programme 209 (FP7/2007-2013) under Grant Agreement No. 601126 210 (HANAS). K.D.J. acknowledges funding from the Marie Sk\l{}odowska Individual Fellowship under REA grant agreement No. 661416 (SiPhoN).

\section*{Acknowledgments}
The authors thank A.~W.~Elshaari, R.~Young, and M.~Stevenson for scientific discussions. K.~D.~J\"ons thanks P.~Michler, P.~Senellart, R.~Trotta, J.~Wildmann, J.~Zhang, T.~Kuroda, G.~Juska, A.~Orieux, S.~Ducci, A.~ Predojevi\'{c}, X.-L. Wang, L.-K. Chen, C.-Y. Lu, J.-W. Pan, K.~Shalm, S.~W.~Nam, R.~Thew, N.~Gisin, M.~Giustina, and A.~Zeilinger for providing the required data for Fig.\,6 and giving valuable feedback. 

 \section*{Author contributions} M.A.M.V., K.D.J, and M.E.R. conceived and designed the experiments. L.S., K.D.J, and M.E.R. performed the experiments. D.D. and P.J.P. fabricated the sample. K.D.J., M.E.R., and L.S. analyzed the data. A.Gu. and A.Gi. developed the detectors. M.E.R and V.Z. supervised the project. K.D.J., L.S., and M.E.R. wrote the manuscript with input from the other authors.

 \section*{Additional information}
Competing financial interests: The authors declare no competing financial interests.




\end{document}